# Analytic Evaluation of Three Different Five – Electron Atomic Integrals Involving Exponentially Correlated Functions of $r_{ij}$ With $r_{ij}$'s Not Forming A Closed Loop

B PADHY

Department of Physics, Khallikote (Autonomous) College,
Brahmapur-760001,Odisha
Email: bholanath.padhy@gmail.com

**Abstract :** The simple method outlined in our earlier paper [ B.Padhy, Orissa Journal of Physics, vol.19, No.1, p.1, February 2012] has been utilized here for analytic evaluation of three different five-electron atomic integrals with integrands involving products of s Slater-type orbitals and exponentially correlated functions of the form $r_{ij} \exp(-\lambda_{ij}r_{ij})$. Only products of those $r_{ij}$'s which do not form a closed loop by themselves, are considered.

## 1. Introduction

One of the most important tasks in quantum theory of many–electron atoms is to carry out calculations for energy and wave function for an atom taking into account the correlation between the positions of the different electrons as discussed in a review article by Lowdin[1]. Neglecting the correlation, the solution of the time independent Schroedinger equation for the wave function of such a system has been obtained by the Hartree-Fock (HF) approximation [2] to a high degree of accuracy. There are two standard variational methods for getting solutions more accurate than the HF solution by including the correlation effects. They are (i) the configuration-interaction (CI) method, and (ii) the Hylleraas (Hy) method. Various CI and Hy studies have been reported in a review article by Silverman and Brigman[3]. The CI method is based on products of one-electron functions. In the Hy method, the interelectronic coordinates $r_{ij}$ are explicitly included in the trial function as was first proposed by Hylleraas[4,5]. Though this method is relatively difficult to apply for calculations in comparison with the CI method, due to the presence of $r_{ij}$ in the trial function, yet it has been observed to be quite successful in case of two-electron atoms[6].

It seems to be fairly difficult ,however, to extend the Hy method to systems with more than two electrons, mainly because the integrals contain several $r_{ij}$'s in the integrands, making the evaluation of such integrals difficult. Even in case of Li atom the integrands contain the product, $(r_{12}r_{23} r_{13})$, of all three interelectronic distances which form a closed loop ( here a triangle ) by themselves. The first attempt of computation of the ground state of Li atom by Hy method was due to James and Coolidge[7], who were successful in expressing the atomic three-electron nine- dimensional integral containing the product $r_{12}r_{23}r_{13}$, in terms of auxiliary functions A,V and W, which are themselves of one-, two-, and three- dimensional integrals, respectively. For definitions and a very thorough knowledge about these auxiliary functions as well as the development of various methods of evaluation, with greater accuracy, of three-electron correlated integrals over spherically symmetric atomic Slater-type orbitals(STO's)[8], one can go through the paper by Pelzl and King[9] as well as the references therein, and the review article by King[10]. The evaluation of more general three-electron correlated integrals over nonspherically symmetric STO's has been reported by Yan and Drake[11]. Even these days, the Hy-type three electron correlated wave functions are being applied for accurate calculation of physical properties of Li and Li- like ions[12-14]. It is very much clear that the computational difficulties associated with extending the Hy method to atoms with more than three electrons are formidable, and hence, there have been no really accurate Hy calculations for atoms with more than three electrons[15].

With the purpose of doing accurate calculations for atoms with more than three electrons, by avoiding computational difficulty met in Hy method, Sims and Hagstrom[15-17] in a series of papers starting from early seventies, utilized the methods of explicit introduction of interelectronic distances $r_{ij}$ into a configuration- interaction wave function, subject to the restriction of at most one $r_{ij}$ per configuration. This Hylleraas-configuration interaction(Hy-CI) variational method has been extensively utilized for atomic calculation over the years. Many of the details of the integrals including two four-electron correlated integrals which arise in an Hy-CI calculation of atomic systems have been discussed by Sims and Hagstrom in their earlier paper[17], wherein one more auxiliary function X was introduced in addition to the auxiliary functions A,V and W. The function X is itself in the form of a four- dimensional integral. Recently, very highly accurate calculations of the auxiliary functions A,V,W and X (redenoted by $A_1, A_2, A_3$ and $A_4$ of the first, second, third and fourth orders, respectively) have been reported in the literature [18-19].

Several reports relating to evaluation of certain four-electron atomic integrals involving Hy type correlated wave functions are available in the literature[17,19-22]. Also Fromm and Hill[23] in 1987 could succeed in obtaining a closed- form expression for the atomic three-electron generating integrals containing the product of all the three interelctronic distances with exponential correlation. Subsequently, four more reports of analytical evaluation relating to the same integral were published[24-26]. Formulas for the recursive generation of many other three-electron exponentially correlated integrals starting from the value of the integral evaluated by Fromm and Hill[23], are recently reported by Harris[27]. Very recently, the author has reported[28] closed-form expressions for certain two-, three-, and four- electron atomic integrals involving product of atomic s STO's and exponentially correlated functions of those $r_{ij}$'s which do not form a closed loop by themselves. The analytic evaluations done by employing Fourier transform technique, are observed to be very simple and straightforward, and do not involve either the spherical harmonic addition theorem or the Feynman technique[29].

As far as the knowledge of the author goes, no report of evaluation of any five-electron atomic integral with exponentially correlated function of $r_{ij}$'s, is available in the literature. Also evaluation of such integrals in closed- form , is likely to help the workers in doing variational calculations for five-electron atoms. With this purpose in mind, the simple method of evaluation outlined earlier by the author, is further extended for obtaining closed-form expressions for three different five-electron atomic integrals involving exponential correlation of the type as proposed.

In what follows, it is assumed that all spin dependences have been factored off, and that the integrations are to be evaluated over the whole of the configuration space of five electrons. It is further assumed that the nucleus is infinitely heavy, and is situated at the origin of the coordinate system.

## 2. Definitions of Integrals

Let us denote the position vector of the $i^{th}$ electron with respect to the nucleus by $\mathbf{r}_i$, and let $\mathbf{r}_i = ( r_i, \theta_i, \phi_i )$ be the spherical polar coordinates of the $i^{th}$ electron , with atomic nucleus as origin. The distance between the $i^{th}$ and the $j^{th}$ electrons is obviously $r_{ij} = | \mathbf{r}_i - \mathbf{r}_j |$. Also the following set of orthonormalized atomic STO's is taken as the orbital basis:

$$u(\mathbf{r}) = A(a,n) \, r^{n-1} \exp(-ar) Y_{l,m}(\theta, \phi),$$

$$A(a,n) = [(2a)^{2n+1} / (2n)!]^{1/2},$$

$$n \geq l+1 \geq 1, \tag{1}$$

where n is the radial quantum number and A(a,n), the normalization constant. The quantum numbers l and m define the order and the degree of the orthonormal spherical harmonic $Y_{l,m}(\theta, \phi)$[30].

The five electron atomic integrals involving exponential correlation are of the form

$$I_t = \int F_1(\mathbf{r}_1) F_2(\mathbf{r}_2) F_3(\mathbf{r}_3) F_4(\mathbf{r}_4) F_5(\mathbf{r}_5) R_t \, d\mathbf{r}_1 d\mathbf{r}_2 \, d\mathbf{r}_3 d\mathbf{r}_4 \, d\mathbf{r}_5 \, ; \, t=1,2,3, \tag{2}$$

where $R_t$ (t=1,2,3) are the correlation factors given by

$$R_1 = ( r_{12} r_{13} r_{14} r_{15})^{-1} \exp(-\lambda_{12} r_{12} - \lambda_{13} r_{13} - \lambda_{14} r_{14} - \lambda_{15} r_{15}), \tag{3}$$

$$R_2 = ( r_{12}\, r_{23}\, r_{34}\, r_{45})^{-1}\, \exp(-\lambda_{12}r_{12}-\lambda_{23}r_{23}-\lambda_{34}r_{34}-\lambda_{45}r_{45}), \tag{4}$$

and

$$R_3 = ( r_{12}\, r_{23}\, r_{34}\, r_{35})^{-1}\, \exp(-\lambda_{12}r_{12}-\lambda_{23}r_{23}-\lambda_{34}r_{34}-\lambda_{35}r_{35}). \tag{5}$$

These are the only possible three different exponentially correlated factors with $r_{ij}$'s not forming a closed loop. The F's are charge distributions given by products of STO's:

$$F(\mathbf{r}) = u^*(\mathbf{r})\, u'(\mathbf{r})$$

$$= A(a,n)A(a',n')\, \exp[-(a+a')r]\, Y^*_{l,m}(\theta, \phi)\, Y'_{l',m'}(\theta, \phi). \tag{6}$$

## 3. Evaluation of Integrals

The evaluation of the $I_t$ integrals corresponding to s STO's is considered in this report. Analytic evaluation involving p,d,f,g STO's and the corresponding discussions are postponed for a future publication, though the work is nearing completion.

The five-electron atomic generating integrals involving s STO's are denoted by $J_1, J_2$ and $J_3$:

$$J_1 = \int d\mathbf{r}_1 d\mathbf{r}_2\, d\mathbf{r}_3 d\mathbf{r}_4\, d\mathbf{r}_5\, (r_1 r_2 r_3 r_4 r_5)^{-1}\, R_1 \exp(-\lambda_1 r_1-\lambda_2 r_2-\lambda_3 r_3-\lambda_4 r_4-\lambda_5 r_5), \tag{7}$$

$$J_2 = \int d\mathbf{r}_1 d\mathbf{r}_2\, d\mathbf{r}_3 d\mathbf{r}_4\, d\mathbf{r}_5\, (r_1 r_5)^{-1}\, R_2 \exp(-\lambda_1 r_1-\lambda_2 r_2-\lambda_3 r_3-\lambda_4 r_4-\lambda_5 r_5), \tag{8}$$

and

$$J_3 = \int d\mathbf{r}_1 d\mathbf{r}_2\, d\mathbf{r}_3 d\mathbf{r}_4\, d\mathbf{r}_5\, (r_1 r_3 r_4 r_5)^{-1}\, R_3 \exp(-\lambda_1 r_1-\lambda_2 r_2-\lambda_3 r_3-\lambda_4 r_4-\lambda_5 r_5), \tag{9}$$

where $R_1, R_2$ and $R_3$ are given by Eqs.(3),(4),(5), respectively. These integrals can be recast as

$$J_1 = \int d\mathbf{r}_1 (r_1)^{-1} \exp(-\lambda_1 r_1) \int d\mathbf{r}_2 (r_2\, r_{12})^{-1} \exp(-\lambda_2 r_2-\lambda_{12}r_{12}) \int d\mathbf{r}_3 (r_3\, r_{13})^{-1} \exp(-\lambda_3 r_3-\lambda_{13}r_{13})$$

$$\times \int d\mathbf{r}_4 (r_4\, r_{14})^{-1} \exp(-\lambda_4 r_4-\lambda_{14}r_{14}) \int d\mathbf{r}_5 (r_5\, r_{15})^{-1} \exp(-\lambda_5 r_5-\lambda_{15}r_{15}), \tag{10}$$

$$J_2 = \int d\mathbf{r}_1 (r_1)^{-1} \exp(-\lambda_1 r_1) \int d\mathbf{r}_2 (r_{12})^{-1} \exp(-\lambda_2 r_2-\lambda_{12}r_{12}) \int d\mathbf{r}_3 (r_{23})^{-1} \exp(-\lambda_3 r_3-\lambda_{23}r_{23})$$

$$\times \int d\mathbf{r}_4 (r_{34})^{-1} \exp(-\lambda_4 r_4-\lambda_{34}r_{34}) \int d\mathbf{r}_5 (r_5\, r_{45})^{-1} \exp(-\lambda_5 r_5-\lambda_{45}r_{45}), \tag{11}$$

and

$$J_3 = \int d\mathbf{r}_2 \exp(-\lambda_2 r_2) \int d\mathbf{r}_1 (r_1 r_{12})^{-1} \exp(-\lambda_1 r_1-\lambda_{12}r_{12}) \int d\mathbf{r}_3 (r_3 r_{23})^{-1} \exp(-\lambda_3 r_3-\lambda_{23}r_{23})$$

$$\times \int d\mathbf{r}_4 (r_4 r_{34})^{-1} \exp(-\lambda_4 r_4-\lambda_{34}r_{34}) \int d\mathbf{r}_5 (r_5\, r_{35})^{-1} \exp(-\lambda_5 r_5-\lambda_{35}r_{35}). \tag{12}$$

All these integrals are to be evaluated from right to left by utilizing the analytical expression derived for the following integral in our earlier paper[28], by employing Fourier transform of $\exp(-\lambda_{ij} r_{ij}) / r_{ij}$ :

$$\int d\mathbf{r}_i (r_i\, r_{ij})^{-1} \exp(-\lambda_i r_i - \lambda_{ij} r_{ij}) = 4\pi\, (\lambda_i^2 - \lambda_{ij}^2)^{-1} r_j^{-1} [\exp(-\lambda_{ij} r_j) - \exp(-\lambda_i r_j)] \tag{13}$$

After lengthy but simple and straightforward algebra, these integrals can be reduced to one-dimensional forms which are evaluated analytically.

### (i) Evaluation of $J_1$

It can be shown that

$$J_1 = B \int_0^\infty dr_1\, G(r_1, \lambda_1, \lambda_2, \lambda_3, \lambda_4, \lambda_5, \lambda_{12}, \lambda_{13}, \lambda_{14}, \lambda_{15}) / (r_1)^3, \tag{14}$$

where

$$B = 1024\pi^5 [(\lambda_2^2 - \lambda_{12}^2)(\lambda_3^2 - \lambda_{13}^2)(\lambda_4^2 - \lambda_{14}^2)(\lambda_5^2 - \lambda_{15}^2)]^{-1},$$

and $G(r_1)$ is a sum of sixteen terms of the form $\exp(-\beta_i r_i)$, $i=1,2,3,\ldots,16$. Each of these sixteen $\beta$'s is a sum of five different $\lambda$'s out of the nine $\lambda$'s in Eq(10), and no two $\beta$'s are identical.

It is easy to show that as $r_1 \to 0$, all the four functions $G(r_1), G_1(r_1), G_2(r_1),$ and $G_3(r_1) \to 0$. Here $G_1$, $G_2$ and $G_3$ represent the first, second and third order differentiation, respectively. Also by employing L'Hospital's rule for 0/0, it can be proved that $G(r_1)/(r_1)^3 \to 0$, $G(r_1)/(r_1)^2 \to 0$, $G_1(r_1)/r_1 \to 0$ and $G_2(r_1)/r_1 \to 0$ as $r_1 \to 0$.

Integrating the right hand side of Eq.(14) by parts, the integral $J_1$ becomes

$$J_1 = \tfrac{1}{2} B \int_0^\infty dr_1\, G_2(r_1) / r_1. \tag{15}$$

Combining the terms in $G_2(r_1)$ suitably with the purpose to make use of the standard integral [31],

$$\int_0^\infty (dx/x)\, [\exp(-ax) - \exp(-bx)] = \ln(b/a), \tag{16}$$

a closed-form expression for $J_1$ is obtained, which is, however, not given here because it is lengthy.

### (ii) Evaluation of $J_2$

It can be shown that

$$J_2 = \int_0^\infty dr_1\, H(r_1, \lambda_1, \lambda_2, \lambda_3, \lambda_4, \lambda_5, \lambda_{12}, \lambda_{23}, \lambda_{34}, \lambda_{45}), \tag{17}$$

where H is a function of $r_1$ and nine different $\lambda$'s. It is a sum of several terms each of which is exponential in natur and hence each term is evaluated to give a gamma function. It is not necessary to use Eq.(16), or L'Hospital's rule for 0/0. The final closed-form expression is lengthy and hence not reported here.

### (iii) Evaluation of $J_3$

$J_3$ is expressed as

$$J_3 = {_0\int^\infty} dr_3 \, [\, D_1 \, f_1(r_3) / (r_3)^2 - D_2 \, f_2(r_3) / (r_3)^2 \,], \tag{18}$$

where

$$D_1 = 1024\pi^5 [(\lambda_4^2 - \lambda_{34}^2)(\lambda_5^2 - \lambda_{35}^2)(\lambda_1^2 - \lambda_{12}^2)\{(\lambda_2 + \lambda_{12})^2 - \lambda_{23}^2\}]^{-1},$$

and

$$D_2 = 1024\pi^5 [(\lambda_4^2 - \lambda_{34}^2)(\lambda_5^2 - \lambda_{35}^2)(\lambda_1^2 - \lambda_{12}^2)\{(\lambda_1 + \lambda_2)^2 - \lambda_{23}^2\}]^{-1}.$$

Both $f_1$ and $f_2$ are functions of $r_3$ and all nine $\lambda$'s that appear in Eq.(12). Each one of $f_1$ and $f_2$ is a sum of eight terms. All these terms are exponential in nature. Employing L'Hospital's rule for 0/0 and the standard integral in Eq.(16), and following the procedure adopted for evaluation of $J_1$, a closed-form expression is obtained for $J_3$, which is lengthy, and hence not reported here.

### 4. Conclusion

It is observed that the method of evaluation of the five-electron atomic integrals involving exponential correlation, as considered here, is simple and straightforward. Closed–form expressions for integrals containing integer powers of $r_i$, $r_{ij}$, etc. can be obtained from the closed-form expressions for these generating integrals by the method of parametric differentiation. It is presumed that the results of this work may help in accurate calculations for five-electron systems.

### Acknowledgement

The author is extremely thankful to the Director, Institute of Physics, Bhubaneswar for providing the library and internet facilities of the Institute during the process of this work, and is grateful to Prof. N.Barik for useful discussions.